\def \SAIT #1 #2 {{\em Mem.\ Soc.\ Astron.\ It.\/} {\bf #1}, #2}
\def \MESS #1 #2 {{\em The Messenger\/} {\bf #1}, #2}
\def \ASTRNACH #1 #2 {{\em Astron. Nach.\/} {\bf #1}, #2}
\def \AAP #1 #2 {{\em Astron. Astrophys.\/} {\bf #1}, #2}
\def \AAL #1 #2 {{\em Astron. Astrophys. Lett.\/} {\bf #1}, L#2}
\def \AAR #1 #2 {{\em Astron. Astrophys. Rev.\/} {\bf #1}, #2}
\def \AAS #1 #2 {{\em Astron. Astrophys. Suppl. Ser.\/
} {\bf #1}, #2}
\def \AJ #1 #2 {{\em Astron. J.\/} {\bf #1}, #2}
\def \ANNREV #1 #2 {{\em Ann. Rev. Astron. Astrophys.\/} {\bf #1}, #2}
\def \APJ #1 #2 {{\em Astrophys. J.\/} {\bf #1}, #2}
\def \APJL #1 #2 {{\em Astrophys. J. Lett.\/} {\bf #1}, L#2}
\def \APJS #1 #2 {{\em Astrophys. J. Suppl.\/} {\bf #1}, #2}
\def \APSS #1 #2 {{\em Astrophys. Space Sci.\/} {\bf #1}, #2}
\def \ASR #1 #2 {{\em Adv. Space Res.\/} {\bf #1}, #2}
\def \BAIC #1 #2 {{\em Bull. Astron. Inst. Czechosl.\/} {\bf #1}, #2}
\def \JSQRT #1
 #2 {{\em J. Quant. Spectrosc. Radiat. Transfer\/} {\bf #1}, #2}
\def \MN #1 #2 {{\em Mon. Not. R. Astr. Soc.\/} {\bf #1}, #2}
\def \MEM #1 #2 {{\em Mem. R. Astr. Soc.\/} {\bf #1}, #2}
\def \PLR #1 #2 {{\em Phys. Lett. Rev.\/} {\bf #1}, #2}
\def \PASJ #1 #2 {{\em Publ. Astron. Soc. Japan\/} {\bf #1}, #2}
\def \PASP #1 #2 {{\em Publ. Astr. Soc. Pacific\/} {\bf #1}, #2}
\def \NAT #1 #2 {{\em Nature\/} {\bf #1}, #2}

\documentstyle{memsait}
\input epsf.sty
\begin{opening}
\title{TYPE II SUPERNOVAE AT HIGH REDSHIFTS} 
\author{N.N.~CHUGAI$^1$, S.I.~BLINNIKOV$^2$, P.~LUNDQVIST$^3$}
\institute{$^1$Institute of Astronomy RAS, Moscow, Russia\\
$^2$Institute for Theoretical and Experimental Physics, Moscow, Russia\\
$^3$ Stockholm Observatory, Saltsj\"{o}baden, Sweden}
\date{} 
\end{opening}
\newcommand{\msun}{$M_{\odot}$}
\newcommand{\rsun}{$R_{\odot}$}

\begin{document}

%\oddpagefooter{\sf Mem. S.A.It., Vol. ??, ??}{}{\thepage}
%\evenpagefooter{\thepage}{}{\sf Mem. S.A.It., Vol. ??, ??}
\oddpagefooter{}{}{} % LEAVE AS IT IS !
\evenpagefooter{}{}{} % LEAVE AS IT IS !

\bigskip

\begin{abstract}
A hydro code coupled with radiation transfer was applied to
produce monochromatic light curves of two models of type II supernovae
(SN~II) simulating SN II-P and SN IIb (SN~1993J-like).
We then used these template light curves to evaluate the possibility
of detecting SNe~II at different redshifts. 
With a 5 hour exposure at VLT/FORS the SN~II-P model
may be detected at $z=1$. However, since our model of SN~II-P is
underluminous at early phase ($t<10$ days) by $\approx 1.5$ mag
a detection at $z=2$ is quite plausible. SN~IIb can
be detected as far as at $z=4$. For 100\% detection efficiency up to $z=2$
one expects to find roughly 1 SN~II yr$^{-1}$ arcmin$^{-2}$.
\end{abstract}

%\tighten

\section{Introduction}

Type II supernovae  trace closely the overall star formation
process and metal enrichment of galaxies and intergalactic space. 
Therefore, the observation of 
high $z$ SNe~II would be a complementary test of  
cosmic evolution models,  being developed intensively during
the latest several years.
The depth of imaging of distant sources reached recently by HST,
Keck I and II, and also by VLT raises an exciting
possibility of observations of SNe~II at $z>1$. 
Due to high UV/optical luminosity after the shock
breakout some of SNe~II may be luminous enough to be detected 
at extremely high $z$, possibly even at pre-galactic epoch $z\geq 10$.
By some reason (e.g., lower metallicity)
at high $z$ we may find also different, possibly even more 
luminous, types of collapse events (e.g., like SN~1998bw).

An obvious prerequisite for high redshift study of SNe~II is a set of 
template rest-frame monochromatic UV/optical
light curves for different subtypes of SNe~II.
Unfortunately,  a complete sample of template observed
light curves is not available at present. Even for the best observed SN~1987A
and SN~1993J, the early epoch following the shock wave breakout is missing.
We note in this context that these supernovae, both representing rare
subtypes, were used for calculations of rate--magnitude relation
by J{\protect \o}rgensen et al. (1997). 

Alternatively, we may rely on theoretical light
curves. However, not until now realistic modeling of SN~II spectra 
has become possible. The hydro-code we have used to do this is
coupled with radiation transfer (Blinnikov and Bartunov 1993,
Blinnikov et al. 1998). It permits us to produce supernova models, which we
then use to recover the early UV/optical behavior of SNe~II. 
The creation of a set of template light curves for different kinds of SNe~II
(viz. SN~II-P, SN~II-L, SN~IIn) requires extensive modeling. This is under
way (Chugai et al. 1999).

Here we report on preliminary results of the evaluation of detection
feasibility for high redshift SNe~II on the basis of two  
models: SN~II-P and a low mass SN~1993J-like event, referred to
as SN~IIb (Woosley et al. 1987). We do not consider
here models of SN~II-L and SN~IIn, which is still a subject of 
study. We estimate also the frequency of SN~II events for 
different modes of cosmic star formation rate and give 
an assessment of SN~II detection in the search with VLT. In this preliminary
study we have not included dust extinction. This will be done in our fuller
analysis (Chugai et al. 1999).

\begin{figure}
\epsfysize=6cm % fix the y-dimension and scales x-dim. to y-dim.
\hspace{3.5cm}\epsfbox{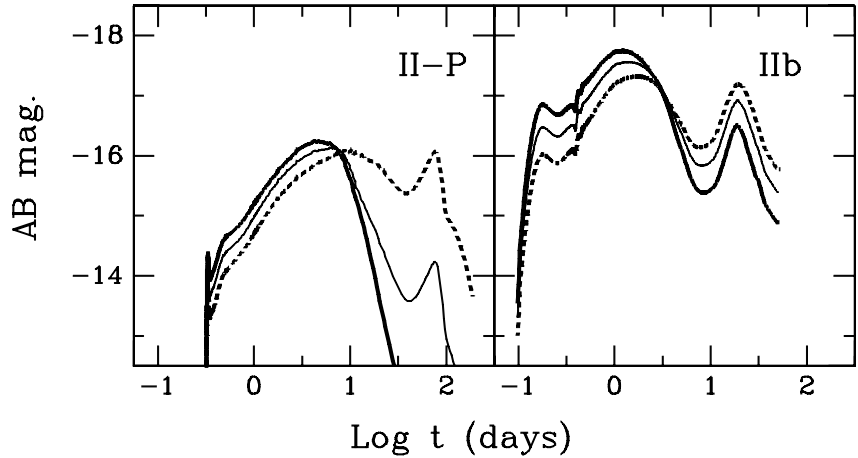} %for centering: act on hspace argument
\caption[h]{Light curves of SN~II models. Absolute monochromatic 
AB magnitude is defined following Oke \& Gunn (1983) with wavelengths 
equal to the effective wavelengths of 
U band (thick solid line), B (thin solid) and V (dotted).}
\end{figure}

\section{Models and monochromatic light curves}

If the wind around pre-SN is of low density, the light curve of 
SN~II is determined only by the ejected mass ($M$), 
kinetic energy ($E$), pre-SN radius ($R_0$), density structure and 
composition of the hydrogen envelope. The bulk of
type II supernovae is SN~II-P. This class is simulated here
by a model with parameters 
$R_0=300$ \rsun, $M=14$ \msun, $E=1.0 \times 10^{51}$ ergs.
The density distribution in the pre-SN is 
that of SN~1987A rescaled to a radius of 300 \rsun.
The underlying model is 14E1 used by Shigeyama \& Nomoto (1990).
The second model represents low mass SN~II or SN~IIb 
with $R_0=300$ \rsun, $M=2.5$ \msun and $E=1.3 \times 10^{51}$ ergs. This is
the model 13C of Woosley et al. (1994) which was used to simulate SN 1993J. 
Although this is a relatively rare subclass, we have
chosen to include a model of the well studied SN~1993J to show the effect of
the relatively strong initial UV/optical 
peak, which is characteristic of low mass extended pre-SN~II. 
In this sense the model SN~IIb may be a good proxy for SN~II-L, 
as well as SN~IIn, which both have rather extended envelopes
presumably related to a strong stellar wind.

The absolute monochromatic light curves calculated
at wavelengths corresponding to the effective
wavelengths in the UBV bands for both models are shown in Fig. 1.
SN~IIb are $\approx 1.8$ mag brighter in UBV than SN~II-P at first
maximum light between 1--10 day. This reflects the different density
structures of the outer layers of these pre-SN models.
It should be emphasized that our SN~II-P model does not represent
an average SN~II-P in detail. Moreover, it is somewhat
underluminous (by $\sim 1.5$ mag) in the UBV bands at the initial phase 
1--10 days compared to observations. For example, SN~1988A, a supernova of
type II-P subclass with a clearly observed early rise, has an initial maximum 
at the level of $M_V=-17.7$ mag (Turatto et al. 1993). This is 1.6 mag brighter 
than in our model ($M_V=-16.1$ mag). The choice of an adequate model 
for SN~II-P requires the reconstruction of the
pre-SN~II-P on the basis of a fit to UBV light curves 
and photospheric velocity evolution of representative SN~II-P. 

\begin{figure}
\epsfysize=8cm % fix the y-dimension and scales x-dim. to y-dim.
\hspace{3.5cm}\epsfbox{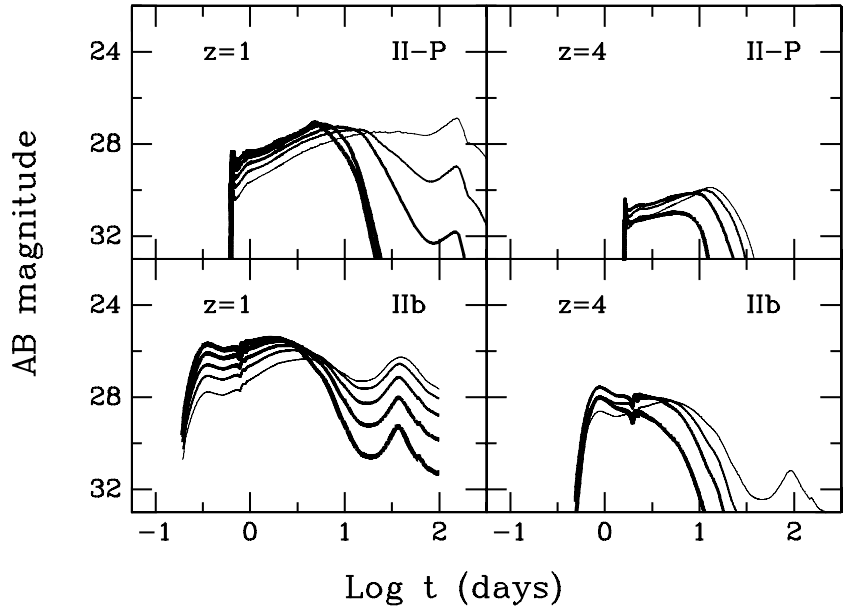} %for centering: act on hspace argument
\caption[h]{Monochromatic BVRIJ light curves for
SN~II models at $z=1$ and $z=4$. From B band (thick line) 
through J band (thin line) line thickness decreases.}
\end{figure}

At redshifts $z=1$ and $z=4$ the models produce the BVRIJ light curves shown
in Fig. 2. The Lyman continuum absorption in the host galaxy and 
the integrated absorption by L$\alpha$ forest have been taken into account,
the latter following the recipe of Press, Rybicky \& Schneider (1993). 
Note that hydrogen absorption is responsible for the suppression 
of the B flux at $z=4$ (Fig. 2). It is remarkable that, while
the light curves are different for different models and bands, 
the maxima of the R, I, and J magnitudes at $z\approx 4$ are quite
similar. Amazingly enough, the light curve of the SN~II-P model in the V band
becomes narrow at high $z$ in spite of time dilation. Such a behavior
stems from the decrease of the rest frame light curve width as 
wavelength becomes shorter.

The most striking property of all the light curves of SN~II
at various $z$ is the fast early rise ($t_{\rm r}\approx 1$ day) 
to maximum (or nearly maximum) luminosity. 
This property, very specific for SNe~II, is
unlike the light curves of  SN~Ia and SN~Ib, which are characterized
by a broad rise to maximum ($t_{\rm r}>20$ days at $z\geq 1$).
The latter is demonstrated by the secondary maximum at $t>30$ days
in the model SN~IIb, which is very similar to the main 
maximum in SN~Ib and SN~Ia models.  
The shape of light curves at the initial phase may be a
crucial signature for the identification of high redshift SN~II events.  

\section{Detectability by VLT and rate of events}

With a 5 hour exposure on VLT/FORS the limiting magnitudes 
in the V, R, and I bands are
27.5 (S/N = 6.2), 27.8 (S/N = 6.7), and 28.2 (S/N = 6.4), respectively,
assuming an airmass of 1.5, and a seeing of 0."8. For such an exposure, 
our model SN~II-P can be detected by VLT 
in the R and I bands at $z\approx 1$ (Fig. 3).
However, since the detection limit of SN~II-P is most likely underestimated
because our SN~II-P model is underluminous by $\sim 1.5$ mag at the initial 
phase $t\geq 10$ days, SN~II-P can probably be seen by VLT at least up to
$z=2$. In the case of SNe IIb a secure detection is guaranteed at $z=2$
and they may be observed even up to $z=4$ in both the R and I bands (Fig. 3). 

\begin{figure}
\epsfysize=8cm % fix the y-dimension and scales x-dim. to y-dim.
\hspace{3.5cm}\epsfbox{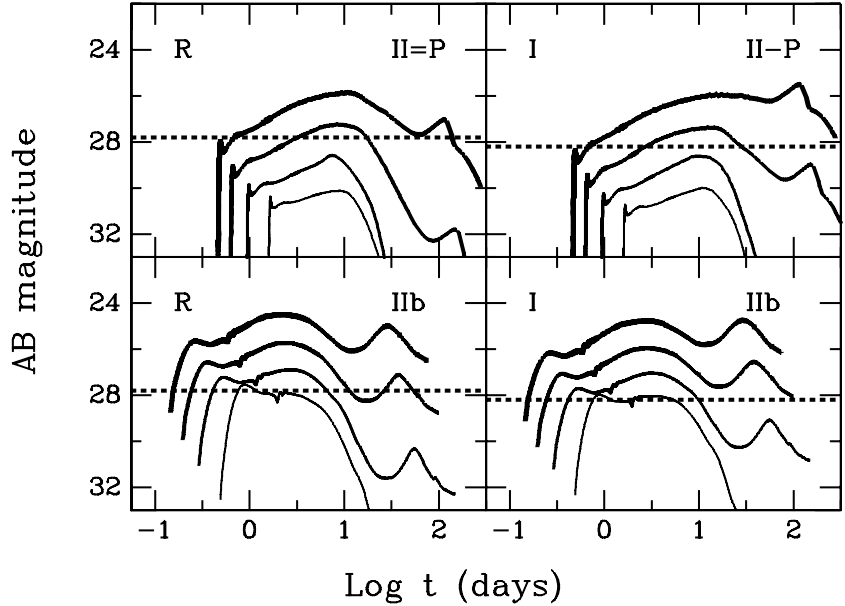} %for centering: act on hspace argument
\caption[h]{Monochromatic R and I light curves of SN~II models at 
different redshifts. Model SN~II-P and model SN~IIb 
light curves are shown for $z$ equal to 0.5, 1, 2, 4 with
line thickness decreasing from $z=0.5$ (thick line) 
through $z=4$ (thin line). The detection limits for 5 hour exposure
at VLT/FORS are shown by dotted lines.}
\end{figure}

There are clear indications that supernovae of subtypes 
SN~II-L (e.g., SN~1979C) and SN~IIn 
(e.g., SN~1994W) produce strong initial UV/optical peak $(M_{B,V} \sim
-(18-19)$. This is related presumably to the presence of an extended 
envelope and/or wind, which are rarefied enough for the shock wave 
energy to escape avoiding severe adiabatic loss. We predict that 
both these kinds of SNe~II could be well observed at redshifts $z\geq 4$. 
However, this claim requires a confirmation in light curve computations.

The observing rate of core-collapse supernovae at some $z$ 
is determined by the co-moving star formation rate (SFR), the initial mass 
function (IMF), the minimum initial
mass of stars producing supernovae, $M_2$, and the cosmological model adopted.
A reasonable estimate of the total rate 
of core collapse supernovae may be obtained assuming 
that all stars with the
present day ratio of stellar density to critical, $\Omega_{\rm s}$, 
were born instantly at some characteristic epoch $z=z_{\rm s}$.
Then adopting flat cosmology (i.e., $\Omega=1$) one gets the observing rate
\begin{equation}
\dot{N}=6c^3\Omega_{\rm s}G^{-1}(dN_{\rm cc}/dm)
[1-1/(1+z_{\rm s})^{1/2}]^2 \approx 60[1-1/(1+z_{\rm s})^{1/2}]^2 
\quad ({\rm s}^{-1}).
\end{equation}
We adopt $\Omega_{\rm s}=0.0042$ (Fukugita, Hogan \& Peebles 1996), and
the number of core collapse supernovae 
per one solar mass converted into stars,
$dN_{\rm cc}/dm \approx 0.012$ \msun$^{-1}$. This value is 
obtained for $M_2=10$\msun, a Salpeter IMF with $\alpha=2.35$, 
$M_{\rm min}=0.3$\msun, $M_{\rm max}=100$ \msun, and a returned (into gas)
mass fraction of $R=0.4$. Assuming the star
formation epoch was $z_{\rm s}=3$, we estimate 15 core collapse supernovae 
per second, or $4.5 \times 10^{8}$ SN yr$^{-1}$. The rate increases when
the star formation epoch is shifted further to the past.

\begin{figure}
\epsfysize=6cm % fix the y-dimension and scales x-dim. to y-dim.
\hspace{3.5cm}\epsfbox{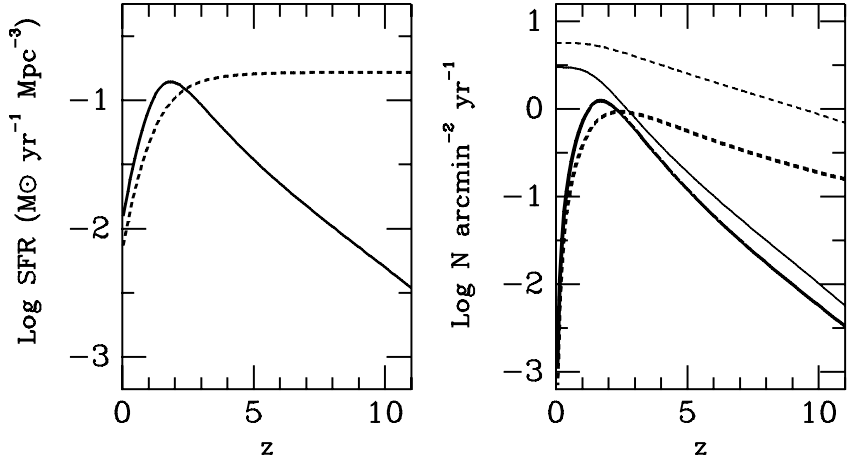} %for centering: act on hspace argument
\caption[h]{Star formation rate and rates of core collapse supernovae.
Left panel shows two types of
star formation evolution: late (solid line) and early (dotted), while 
on the right panel shown are corresponding ``observed'' rate of supernovae 
per unit of $z$ (thick lines) and cumulative rate (thin lines).} 
\end{figure}

For the more general case of flat cosmology 
($\Omega_{\rm M}+ \Omega_{\rm \Lambda}=1$)
with $\Omega_{\rm M}=0.3$, $H_0=65$ km s$^{-1}$ Mpc$^{-1}$
and two different SFR evolution laws (late and early star formation
starting at $z=18$, Fig. 4) the rate per unit of $z$ and 
cumulative rate $N(>z)$ is
shown in Fig. 4. The total cumulative rate is
$\approx 3$ SN yr$^{-1}$ arcmin$^{-2}$ and 
$\approx 5.6$ SN yr$^{-1}$ arcmin$^{-2}$ for late and early star
formation, respectively, while medians of the rate vs. redshift 
distribution are 2.7 and 4.5, respectively.  

Keeping in mind that SNe~II contribute 80\% 
to all core collapse supernova events, we then expect to observe
roughly  $\approx 1$ SN~II  yr$^{-1}$ arcmin$^{-2}$ at $z\leq 2$
provided the detection efficiency is close to 100\%.

The search and identification strategy for SNe~II with typical maximum 
magnitudes around 26-27 should exploit specific shapes of light curves and 
color evolution of SNe~II at high $z$ in V, R, and I.
An interesting approach may be based on the sudden appearance
at time scale $\sim 1$ day. This prompts an observing strategy 
for SN~II detection using two images separated by the time interval
($\Delta t$),
which must be short enough to catch SNe~II using its fast 
appearance, and
yet long enough to pick up as large as possible number of events, 
say  $\Delta t \approx 10$ days. Given two deep exposures separated
by 10 days and assuming the rate 1 SN~II yr$^{-1}$ arcmin$^{-2}$ 
we should then find on average 1.3 SNe~II in the field 
of view of FORS ($6.8'\times 6.8'$). In the case of VIMOS
with the field $14'\times 14'$ we expect to detect in a similar 
observing situation around 5 SNe~II.

\acknowledgements{
We thank Ken Nomoto and Stan Woosley for the presupernova models
and for support. This
research was also supported by the Russian Foundation for Fundamental Research
(96-02-19756 and 98-02-16404), the International 
Science \& Technology Center 97-370, and by
the Royal Swedish Academy of Sciences. P.L. acknowledges further support from 
the Swedish Natural Science Research Council.
}

\end{document}